\documentclass[pdflatex,sn-mathphys-num]{sn-jnl}

\usepackage{graphicx}%
\usepackage{multirow}%
\usepackage{amsmath,amssymb,amsfonts}%
\usepackage{amsthm}%
\usepackage{mathrsfs}%
\usepackage[title]{appendix}%
\usepackage{xcolor}%
\usepackage{textcomp}%
\usepackage{manyfoot}%
\usepackage{booktabs}%
\usepackage{algorithm}%
\usepackage{algorithmicx}%
\usepackage{algpseudocode}%
\usepackage{listings}%
\usepackage{tikz}
\usepackage{multicol}
\usepackage{cases}

\theoremstyle{thmstyleone}%
\theoremstyle{thmstyletwo}%

\theoremstyle{thmstylethree}%

\DeclareUnicodeCharacter{2212}{-}
\begin{document}

\title[Article Title]{Aharonov-Bohm Scattering From Knots}


\author{Kaustav Giri\footnote{kaustavg@cmi.ac.in}}

\author{~V.V. Sreedhar\footnote{sreedhar@cmi.ac.in}}

\affil{\orgname{Chennai Mathematical Institute}

\orgaddress{\street{Plot H1,~SIPCOT IT PARK, Siruseri} 

\city{Chennai}, \postcode{603103},

\country{India}}}


\abstract{The celebrated Aharonov-Bohm effect is perhaps the first example in which the the interplay between classical topology and quantum theory was explored. This connection has continued to shed light on diverse areas of physics like quantum statistics, anomalies, condensed matter physics, and gauge theories. Several attempts were made to generalize the Aharonov-Bohm effect by modifying the simple solenoidal current distribution used by them to the case of multiple solenoids, and a toroidal solenoid, for example. A particularly ambitious task is to confine the magnetic flux to the interior of a knotted solenoid. While it is to be expected that a non-trivial phase factor will be picked up by the wave function of a charged particle travelling in the complement of the knot in three-dimensional space, the lack of symmetry defied attempts to explicitly solve the associated scattering problem. In this paper we report on a way to make progress towards this problem based on multipole expansions. The vector potential produced by a knot is obtained by making a multipole expansion, which is then used to calculate the S-matrix for the scattering of the charged particle, in the Born approximation. It is found that the S-matrix carries an imprint of the knottedness at the octopole order. For the case of a torus knot, a curious factorization property is seen to hold.}




\maketitle

\section{Introduction}\label{sec1}

It is well known that charged particles moving through force-free regions can get diffracted or scattered in quantum mechanics. This was first established by Ehrenberg and Siday \cite{Ehrenberg1949}, and independently by Aharonov and Bohm \cite{Aharonov1959}, who were trying to understand the physical significance of  electromagnetic potentials in quantum mechanics. To establish their results, they had considered the motion of a beam of charged particles in the presence of an impenetrable, infinitesimally thin, and infinitely long solenoid which encloses a uniform magnetic field.
\\

The scalar and vector potentials appearing in the Hamiltonian of the charged particle in the presence of an electromagnetic field are just mathematical artefacts. The classical  motion of the particle, as determined by the Euler-Lagrange or 
Hamilton's equations, is entirely described by the electric and magnetic fields through the well-known Lorentz force law,
$\Vec{F}=q[\Vec{E}+\Vec{v}\times \Vec{B}]$. Hence,
a particle moving through a field-free region, even if the potentials are non-vanishing, remains unaffected. 
\\

In quantum mechanics, however, the wave function of 
a particle moving through a field-free region gets affected by a non-zero vector potential by picking 
up a path-dependent phase factor which produces interference patterns in non-simply connected spaces
as in the aforementioned solenoidal configuration. 
The wave function of the particle, obtained by solving the corresponding Schrodinger equation, clearly shows the deviation from the direction of 
the incident beam.
\\

An alternative insight into the problem is obtained by using the path integral approach. A direct application of the Feynman prescription is not adequate in this case since the classical paths fall into different homotopy classes because of the topological obstruction created by the impenetrability of the solenoid. As shown by Schulman \cite{Schulman1968}, one needs to perform a path integration in the usual way in each equivalence class, and then perform a sum over equivalence classes appropriately weighted by the representations of the fundamental group of the non-simply connected space. This generalization of the path integral approach to quantizing systems on non-simply connected spaces was used by Gerry and 
Singh \cite{Gerry1979} to essentially reproduce the results of Aharonov and Bohm. 
\\

An equally insightful observation by Berry consists in taking account of the non-trivial topology directly at the quantum level by attaching appropriate Dirac phase factors to so-called 
whirling-wave components of the total wave function before performing a resummation to get a 
single-valued wave function \cite{Berry1980}. 
\\

Stovicek \cite{stovicek1989green} developed a technique to study this problem on the universal covering space which allowed him to calculate the Green's function, both for 
the single and two-solenoid cases. 
\\

A rigorous analysis of the problem was performed using
formal scattering theory methods by Ruijsenaars \cite{RUIJSENAARS19831} with particular emphasis on clearing  objections regarding the continuity of the wave function across the walls of the solenoid and the leakage of the magnetic field through 
edge-effects, owing to the physical impossibility of having an infinitely long solenoid.
\\

In a closely-related work designed to silence 
the latter criticism Peskin and Tonomura 
\cite{Peshkin1990} reaffirmed the findings of Aharonov and Bohm by replacing the infinitely long solenoid with a finite toroidal winding which produces 
a magnetic field with support only inside the torus. 
\\

In this paper we consider the scattering of a charged 
particle from an infinitesimally thin knotted solenoid. The magnetic field
is confined to an infinitesimally thin tube of uniform cross-section around the knot and is tangential to the knot at each point. Clearly, since the vector potential produced by such a knotted solenoid is non-vanishing outside the knot even though the magnetic field is confined within the knot \cite{Sreedhar2014}\cite{Sreedhar:2013ca}, and since different paths of the charged particle can wind around the knot differently \cite{buniy2009aharonov}, one expects the charged particle to be scattered. The question we pose is the following: How does the scattering cross-section, or the the S-matrix, carry the imprint of the knottedness of the solenoid?
\\

At first sight, the problem doesn't seem difficult.
All one needs to do is substitute the expression 
for the vector potential produced by a knotted 
solenoid, which can be derived in a straightforward 
manner by using the Biot-Savart law, in the Schrodinger equation and solve it to get the wave function. It 
turns out that this is easier said than done. For example, the $(p,q)$ torus knots -- a class of knots of which the trefoil is the simplest -- are obtained by considering closed loops on the surface of a torus which wind around the two inequivalent cycles of the torus, $p$ and $q$ number of times, $p$ and $q$ being relatively prime integers. The trefoil corresponds to a (2,3) or (3,2) torus knot. The natural choice of orthogonal coordinates to write the Schrodinger equation for such knots is the toroidal coordinate system. However, the Schrodinger equation is not separable in these coordinates \cite{Morse1954}. Moreover there is
no obvious way of prescribing the boundary conditions that make the wave function vanish at the surface of the knotted solenoid.
\\

One can use the path-integral method {\it a la} Gerry 
and Singh, or the whirling-wave approach due to Berry to attach phase factors, given by 
an appropriate representation of the 
knot group, which is the fundamental group of the 
knot complement -- the relevant non-simply connected 
space in this case. This makes the calculations  considerably more
complicated than the corresponding calculations in the standard  Aharonov-Bohm configuration. \\

In principle, we can also use the Green's function method of Stovicek, but this would also require prescribing suitable boundary conditions at the surface of the knotted solenoid in solving the differential equation, as 
already mentioned.
\\

The main issue, as is evident from the above comments, is that a knot is an embedding of a circle in three-dimensional space and doesn't have any obvious symmetry, unlike the infinite straight solenoid, or the toroidal
solenoid. In this paper, we navigate through this problem by the following strategy: We model the knotted solenoid by a collection of point magnetic dipoles, one  at each point on the knot pointing along the tangent to the knot at the point. The magnetic vector potential produced by such a dipole distribution can be calculated from the general expression given by the Biot-Savart integral, by employing the multipole expansion. For calculating the scattering 
amplitude, we find it convenient to enclose the knotted solenoid under consideration by an imaginary minimal sphere. The charged particle, restricted to move outside the 
sphere, experiences the vector potential produced by the knotted solenoid. It may be recalled that a similar trick was used by Ruijsenaaars \cite{RUIJSENAARS19831}, but for different reasons. We calculate the scattering amplitude due to the knotted solenoid in the Born approximation which is the central result of this paper. Some unexpected factorization properties between the scattering amplitudes of knotted and unknotted solenoids are unravelled in the process, in the case of torus knots.
\\

The rest of this paper is organised as follows: In the next section, we compute the magnetic vector potential due to the dipole distribution along the knot using the multipole expansion method. It turns out 
that in order to capture the knottedness of the solenoid, we need to consider not
just the quadrupole moment, which is usually associated with the shape of the distribution, but also the octopole moment. We note some interesting patterns 
satisfied by the multipole moments. In section III, we use the vector potential thus computed to calculate 
the scattering amplitudes for a charged free particle travelling in the complement of the knot. In section IV we calculate the S-matrix for the scattering of a charged particle from a $(p,q)$ torus knot. A curious factorization property associated with this S-matrix is pointed out. In section V, we summarise the results and present 
an outlook. The Appendices contain some
calculational details. 
\\

\section{The Multipole Expansion of the Vector Potential of a Knot}\label{sec2}
As explained in the previous section, we 
model the knot by a uniform distribution of point magnetic dipoles, one at each point of the knot, pointing in the direction of the tangent to the knot at that point. The vector potential produced by the knot is obtained by integrating over the contributions due to individual dipoles along the knot.  

The magnetic vector potential due to a point dipole $\vec{m}$ located at $\vec{r}'$ in three dimensional space has the following form in the Coulomb gauge, $\vec\nabla\cdot \vec A =0$
\begin{equation}
  \vec{A}_{\vec{m}}(\vec{r})\: = \: \frac{\mu_0}{4\pi} \frac{\vec{m}(\vec{r}') \times (\vec{r}-\vec{r}')}{|\vec{r}-\vec{r}'|^3}  
\end{equation}
Let $d\vec{m}(\vec{r}')$ denote the total dipole moment contained in an infinitesimal length $dl$ of the knot $K$, at the position 
$\vec{r}'$. Integrating this over $K$ gives,

\begin{equation}
    \vec{A}_{K}(\vec{r})=\frac{\mu_0}{4\pi} \int_{K} \frac{d\vec{m}(\vec{r}') \times (\vec{r}-\vec{r}')}{|\vec{r}-\vec{r'}|^3}
\end{equation}
\\
At this juncture, it is useful to introduce a parameter $0\leq \tau \leq 2\pi$ along the knot.The unit tangent vector at any point on the knot is given by $\hat t = {d\vec r'\over d\tau}/ \mid{d\vec r^{'}\over d\tau}\mid$. It follows that $d\vec{m}(\vec{r}') = \mathcal{M} dl\hat t  
 = \mathcal{M} \mid{d \vec{r}'\over d\tau}\mid d\tau \hat t$ where, $\mathcal{M}$ is the magnitude of the dipole moment density which we assume to be  uniform. Therefore,
$d\vec {m} (\vec r') = \mathcal{M} \frac{d\vec{r'}}{d\tau} d\tau$ and
\\
\begin{multline}
\label{3}
    \vec{A}(\vec{r})=\frac{\mu_0 \mathcal{M}}{4\pi}\int_{K} d\tau\frac{\frac{d\vec{r}'(\tau)}{d\tau} \times (\vec{r}-\vec{r}'(\tau))}{r^3} ~~\times \\
    \sum_{m,n,k=0}^{\infty} \left(\frac{r'(\tau)}{r} \right)^{m+n+k} P_m(cos\eta(\tau)) P_n(cos\eta(\tau)) P_k(cos\eta(\tau))
    \end{multline}
where we dropped the subscript $K$ on $\vec A$ for the sake of simplicity and  used the standard expression, 
\begin{equation}
    \frac{1}{|\vec{r}-\vec{r}'|}=\frac{1}{r}\sum_{m=0}^{\infty} \left(\frac{r'}{r} \right)^m P_m(cos\eta)
\end{equation}
for $|\Vec{r}|>>|\Vec{r'}|$, 
$\eta$ being the angle between $\Vec{r}$ and $\Vec{r}'$.
\\
The triple sum in the equation \eqref{3} can be expanded in powers of ${r'\over r}$ to get
$\left(1+3(\frac{r'}{r})cos\eta+\frac{3}{2}(\frac{r'}{r})^2(3cos^2\eta-1)+3(\frac{r'}{r})^2cos^2\eta+ \cdots \right)$, where we have used
the well-known expressions for the first three Legendre polynomial functions, {\it i.e,} $P_0({\hbox{cos}}\eta)=1,P_1({\hbox{cos}}\eta)={\hbox{cos}}\eta$ and $P_2(cos\eta)=\frac{1}{2}(3 cos^2\eta-1).$
Therefore,
\begin{multline}
    \vec{A}(\vec{r})=\frac{\mu_0 \mathcal{M}}{4\pi}\int_{k} d\tau \frac{\frac{d\vec{r}'(\tau)}{dt} \times (\vec{r}-\vec{r}'(\tau))}{r^3}~~\times\\
    \left(1+3(\frac{r'}{r})cos\eta+\frac{3}{2}(\frac{r'}{r})^2(3cos^2\eta-1)+
    3(\frac{r'}{r})^2cos^2\eta+\dots  \right)
    \end{multline}
It is now straightforward to calculate the vector potential due to the knot to the desired order in the multipole expansion. 

\subsection{Dipole Moment}
It turns out that the vector potential vanishes to the order $({1\over r^2})$. This is 
not surprising: Since the knot is 
a closed loop, and the dipoles are aligned along the knot, if we imagine a dipole to be a tiny magnet, the north pole of any such magnet is 
coincident with the south pole of the next magnet, effectively cancelling the total dipole moment of the knot.

\subsection{Quadrupole Moment}
To the order $({1\over r^3})$, the vector potential,denoted by $\Vec{A_{[3]}}(\Vec{r})$, is given by 
\begin{equation}
    \vec{A}_{[3]}(\vec{r})=\frac{\mu_0 \mathcal{M}}{4\pi r^3}\int_{K}d\tau ~\Bigl\{3\left(\frac{d\Vec{r}'(\tau)}{d \tau} \times \hat{r}
 \right) \times (r' \hbox{cos}\eta(\tau))-\frac{d\Vec{r}'(\tau)}{d \tau} \times \Vec{r}'(\tau) \Bigl\}
 \end{equation}
At this stage, we choose Cartesian coordinates to describe the knot, namely $\vec r'(\tau) = x'(\tau)\hat i + y'(\tau)\hat j + z'(\tau)\hat k$. 
On the other hand, it turns out to be convenient to use spherical polar coordinates to describe a generic point in three-dimensional space for calculating the quantum mechanical scattering amplitudes in the next section.  
\\

The components of the vector potential to this order can be computed explicitly by integrating over the knot and can be compactly represented by the following expression, 
\\
\begin{equation}
A_{[3]}^{r_i}(r,\theta,\phi)=\frac{\mu_0 \mathcal{M}}{4 \pi r^3}\left[\sum_{j \leq k=1}^{3} 3 Q^{r_i}_{R_jR_k} R_{j} R_{k}~+Q^{r_i}\right]
\end{equation}
\\
where, $i,j,k = 1,2,3$, $r_1 =x,~ r_2 = y,~ r_3 = z$ and $R_1 = \alpha={\hbox{sin}}\theta{\hbox{cos}}\phi,~R_2 = \beta={\hbox{sin}}\theta{\hbox{sin}}\phi, ~R_3 = \gamma={\hbox{cos}}\theta.$
\\
The coefficients $Q^{r_i}_{R_jR_k}$,which are symmetric under the exchange of the lower two indices, can be represented in terms of three numbers $K^i$ as follows:
\\
\begin{numcases}
{Q^{r_i}_{R_j R_k}=}
    0 & if $i \neq j \neq k$ \label{8}
    \\
    \delta_{jk}K^{i}\left(1-\delta_{ij} \right)-\delta_{ik} K^{j}\left(1-\delta_{ij}\right)-\delta_{ij}K^{k}\left(1-\delta_{ik}\right) & otherwise \label{9}
\end{numcases}
\\
where it should be noted that we are {\it not} using Einstein summation convention, and  
\begin{equation}
\label{10}
K^{1}=\int_{K} \left(z'\frac{dy'}{d \tau}\right) d\tau
\end{equation}
\begin{equation}
\label{11}
    K^{2}=\int_{K} \left(x' \frac{dz'}{d\tau}\right) d\tau
    \end{equation}
    \begin{equation}
    \label{12}
    K^{3}=\int_{K} \left(y' \frac{dx'}{d\tau}\right) d\tau.
\end{equation}
\\
Finally, 
\begin{equation}
\label{13}
    Q^{r_i} = -2K^i
\end{equation}
is easily identified to be minus the trace of $Q^{r_i}_{R_jR_k}$.
\\

\subsection{Octopole Moment}
To the order of $\left(\frac{1}{r^4}\right)$, the magnetic vector potential
   $ \vec{A}_{[4]}(\Vec{r})$
    is given by the expression
    \begin{equation}
    \frac{\mu_0 \mathcal{M}} {4 \pi r^4} \int_{K} d\tau \left\{ \frac{3}{2} \left(\frac{d \Vec{r}'(\tau)}{d \tau} \times \hat{r}\right) \times r'(\tau)^2 (5 \hbox{cos}^2\eta-1)-3\left(\frac{d\Vec{r}'(\tau)}{d \tau} \times \Vec{r}'\right)\times (r' \hbox{cos} \eta) \right\}
\end{equation}
The components of the vector potential can be explicitly calculated to be
\begin{equation}
    A_{[4]}^{r_i}(r,\theta,\phi)=\frac{\mu_0 \mathcal{M}}{4 \pi r^4}\int_{K} d\tau \left\{\frac{15}{2}\sum_{j\le k \le l =1}^{3} O^{r_i}_{R_j R_k R_l} R_j R_k R_l-\frac{3}{2}\sum_{p=1}^{3}  O^{r_i}_{R_p} R_p \right\}
\end{equation}
where, $i,j,k,l,p=1,2,3$ and $R_1=\alpha=\hbox{sin}\theta \hbox{cos}\phi,R_2=\beta=\hbox{sin} \theta \hbox{sin} \phi,R_3=\gamma=\hbox{cos} \theta$ as before. In this notation, the tensors $O^{r_i}_{R_jR_kR_l}$ appearing in the octopole moments can be written in the following form,
\\

\begin{equation}
\label{16}
O^{r_i}_{R_j R_k R_l}
\end{equation}
\[   \left\{
\begin{array}{ll}
      =\sum_{m=1}^{3}\epsilon^{jim} r_j r_j \frac{dr_{m}}{d\tau} ~~~\hfill{\hbox{for}}~ j=k=l \\
      =\delta_{jk}\left[\epsilon^{kli} r_k r_k \frac{dr_{k}}{d\tau}-2 \epsilon^{ikl} r_{k}r_{l}\frac{dr_{l}}{d\tau}\right]+\delta_{kl}\left[2\epsilon^{ijl}r_jr_l\frac{dr_{j}}{d\tau}-\epsilon^{jki}r_kr_k\frac{dr_{k}}{d\tau}\right] \\
      ~~~\hfill{\hbox{for}}~ i\neq j,i \neq k, i \neq l , \delta_{jk} \delta_{kl}=0 \\
      =\sum_{m=1}^{3} 2 \left[\delta_{il} \epsilon^{jlm} r_j r_l \frac{dr_{m}}{d \tau}-\delta_{ij} \epsilon^{ikm} r_i r_k \frac{dr_{m}}{d \tau} \right] \\~~\hfill{\hbox{for}} ~[i=j,\delta_{kl}=1] ~{\hbox{or}}~ [i=l,\delta_{jk}=1],\delta_{jk}\delta_{kl}=0 \\
      =\sum_{m=1}^{3} r_k r_k \left[\delta_{kl} \epsilon^{jkm}\frac{dr_{m}}{d \tau}-\delta_{jk} \epsilon^{jlm} \frac{dr_{m}}{d \tau} \right] \\
      ~~\hfill{\hbox{for}}~i=k , \delta_{jk} \delta_{kl}=0 ,\delta_{jk}=1~ {\hbox{or}}~ \delta_{kl}=1\\
      =\sum_{m,n=1}^{3} 2  \epsilon^{imn} r_i r_m \frac{dr_{m}}{d \tau}~~\hfill{\hbox{for}}~ j \neq k \neq l \\
\end{array} 
\right. \]
\\
and 
\begin{equation}
\label{17}
    O^{r_i}_{R_p}=\sum_{m=1}^{3} \left(1+2 \delta_{pm} \right)O^{r_i}_{R_p R_m R_m}
\end{equation}
Note that $r_1=x, r_2=y, r_3=z$, and, like in the quadrupole moment, we are {\it not}  using Einstein's summation convention.  
\\
The octopole moments are totally symmetric under the exchange of all three lower indices, {\it i.e},
$O^{r_i}_{R_j R_k R_l}=O^{r_i}_{R_{\sigma(j)} R_{\sigma(k)} R_{\sigma(l)}}$; $\sigma$ representing all possible permutations among $j,k,l$.
\\

Putting everything together, we get the following expression for the components of the vector potential, of a knotted solenoid upto the order of an octopole moment in the multipole expansion,
\begin{multline}
\label{18}
    A^{r_i}=\frac{\mu_0 \mathcal{M}}{4 \pi r^3}\left(\sum_{j\le k=1}^{3} 3~Q^{r_i}_{R_j R_k}R_j R_k+Q^{r_i}\right)+\\
    \frac{\mu_0 \mathcal{M}}{4 \pi r^4}\left(\frac{15}{2}\sum_{j\le k \le l=1}^{3} R_j R_k R_l\int_{K} d\tau O^{r_i}_{R_j R_k R_l}-\frac{3}{2}\sum_{p=1}^{3}R_p \int_{K} d\tau O^{r_i}_{R_P}\right)
\end{multline}
where,~$\Vec{A}=A^x \hat{i}+A^y \hat{j}+A^z \hat{k}$
\\
\section{The S-Matrix}\label{sec3}
In this section, we calculate the S-matrix for the scattering of a charged particle from a knotted flux tube. As is well-known, the Hamiltonian for such a particle is given by
\\
\begin{equation}
    H=\frac{1}{2M}\left(\Vec{P}-\frac{e\Vec{A}}{c}\right)^2=H_0+V
\end{equation}
where $\vec A$ is the vector potential produced by the knotted flux tube and experienced by the particle of mass $M$, charge $e$, and momentum $\vec P$.
\\
Here $H_0 = \frac {\vec P^2}{2M}$ is the unperturbed Hamiltonian, and $V = -\frac {e}{2Mc}(\vec A\cdot\vec P+\Vec{P}\cdot \Vec{A})$ is the 
interaction which depends on the vector potential $\vec A$, computed to order ${\cal O}\bigl({1\over r^4}\bigr)$ in the previous section. We drop the $A^2$ term which is of ${\cal O}\bigl({1\over r^6}\bigr)$ and lower.

In the scattering problem, we are interested in finding the transition amplitude of a particle from one  eigenstate of the unperturbed Hamiltonian $H_0$ to another.i.e,
\begin{equation}
    <n|\hat{U}_{I}(t,t_0)|i>
\end{equation}
Here $|i>$ and $|n>$ are the initial and final states of the particle and $\hat{U}_{I}$ is the unitary time evolution operator in the interaction picture, given by
\\
\begin{equation}
    \hat{U}_{I}(t,t_0)=1-\frac{i}{\hbar} \int_{t_0}^{t} \hat{V}_{I}(t') \hat{U}_{I}(t',t_0) dt'
\end{equation}
where $\hat{V}_I(t) = e^{\frac{i\hat{H}_0t}{\hbar}} \hat{V}
e^{\frac{-i\hat{H}_0t}{\hbar}}$ is the potential in the interaction picture. In the first order approximation, {\it i.e}, putting $\hat{U}_{I}(t',t_0)= {\mathbf 1}$, the matrix elements are,
\begin{equation}
    <n|\hat{U}_{I}(t,t_0)|i>=\delta_{ni}-\frac{i}{\hbar}V_{ni} \int_{t_0}^{t} e^{iW_{ni} t'} dt'
\end{equation}
where we use the obvious notation, $V_{ni}=<n|\hat{V}|i>$ and $\hbar W_{ni}=(E_n-E_i).$
\\

In the present problem, the potential due to the knot is nonvanishing everywhere. Hence, to get the $S$-matrix elements we simply extend  the limits of integration from $-\infty$ to $\infty$ to get,
\begin{equation}
    S_{ni}~=~<n|\hat{U}_{I}(\infty,-\infty)|i>~=~\delta_{ni}-2\pi i \delta(E_n-E_i)V_{ni}
\end{equation}
Note that at $t=\pm \infty$, the particle is infinitely far away from the knot, the multipole expansion dies off as an inverse power of the distance, the potential vanishes, and we may use free particle eigenstates for $|i>$ and $|n>$.
\\

\subsection{Strategy for Calculating $V_{ni}$}
$V_{ni}$ is given by 
\begin{equation}
    V_{ni}=-\frac{e}{2Mc}\int \psi_{n}^*(\Vec{r})\left(\hat{\Vec{A}}.\hat{\Vec{P}}+\hat{\Vec{P}}.\hat{\Vec{A}}\right)\psi_{i}(\Vec{r})~dV
\end{equation}
where the integration is taken over the complement in three-dimensional Euclidean space, of a ball $B^{3}$, of radius $\lambda_0$, enclosing the knot. Putting~$\psi_i(\Vec{r})=e^{i\Vec{k_i}.\Vec{r}}$ and $\psi_n(\Vec{r})=e^{i\Vec{k_n}.\Vec{r}}$ for the  initial and final energy eigenstates in the position representation, with momentum vectors $\Vec{k}_i$ and $\Vec{k}_n$ respectively, we get,
\begin{equation}
\label{25}
V_{ni}=-\frac{e}{2Mc}\int_{V}  e^{i(\Vec{k_i}-\Vec{k_n}).\Vec{r}} \left[\sum_{r_i=x,y,z} A^{r_i}(\Vec{r})(k_i+k_n)_{r_i} \right] dV
\end{equation}
\\
In writing the above equation, and henceforth, we set $\hbar =1$ to avoid clutter. 

From equation \eqref{18}, it follows 
\\

$\sum_{r_i=x,y,z} A^{r_i}(\Vec{r})(k_i+k_n)_{r_i}$
\begin{multline}
    =\sum_{r_i=x,y,z}\frac{\mu_0 \mathcal{M}}{4 \pi r^3}\left(\sum_{j\le k=1}^{3} 3~(k_{i}+k_n)_{r_i}Q^{r_i}_{R_j R_k}R_j R_k+(k_{i}+k_n)_{r_i}Q^{r_i}\right)+\\
    \sum_{r_i=x,y,z}\frac{\mu_0 \mathcal{M}}{4 \pi r^4}(k_{i}+k_n)_{r_i}\times \left(\frac{15}{2}\sum_{j\le k \le l=1}^{3} R_j R_k R_l\int_{K} d\tau~O^{r_i}_{R_j R_k R_l}-\frac{3}{2}\sum_{p=1}^{3}R_p \int_{K} d\tau~O^{r_i}_{R_P}\right)
\end{multline}
Substituting the above expression in equation \eqref{25} and using the standard expression of the plane wave, {\it viz.} 
\\

\begin{equation}
    e^{i(\vec{k}.\vec{r})}=4\pi\sum_{l=0}^{\infty}\sum_{m=-l}^{m=l}i^l j_{l}(kr)Y_{lm}(\hat{k})Y^{*}_{lm}(\theta,\phi)
\end{equation}
where,~$Y_{lm}(\theta,\phi)$ are spherical harmonics,~$j_l(r)$ are spherical Bessel functions and the momentum change $\Vec{k}=\Vec{k}_i-\vec{k}_n$. We get the matrix element $V_{ni}$ to be 

\begin{multline}
\label{28}
  V_{ni}=\\
  -\frac{\mu_0 \mathcal{M} e}{2Mc}\sum_{j \le k=1}^{3}\left\{[\sum_{r_i=x,y,z}3~(k_i+k_n)_{r_i} Q^{r_i}_{R_j R_k}] \int \sum_{l,m}A(l,m) Y^{*}_{lm}(\theta,\phi)R_j R_k d\Omega \right\} \\
   -\frac{\mu_0 \mathcal{M} e}{2Mc}[\sum_{r_i=x,y,z}(k_i+k_n)_{r_i}Q^{r_i}] \left\{\sum_{l,m} A(l,m) \int Y^{*}_{lm}(\theta,\phi) d\Omega \right\} \\
   -\frac{15}{2}\frac{\mu_0 \mathcal{M} e}{2Mc}\sum_{j \le k \le l=1}^{3}\left\{[\sum_{r_i=x,y,z}(k_i+k_n)_{r_i} \int_{K} d\tau O^{r_i}_{R_j R_k R_l}] \int \sum_{l,m}B(l,m) Y^{*}_{lm}(\theta,\phi)R_j R_k R_l d\Omega \right\}\\
   +\frac{3}{2}\frac{\mu_0 \mathcal{M} e}{2Mc} \sum_{p=1}^{3}\left\{[\sum_{r_i=x,y,z} (k_i+k_n)_{r_i} \int _{K} d \tau O^{r_i}_{R_P}] \int \sum_{l,m} B(l,m) Y^{*}_{lm}(\theta, \phi) R_{P} d\Omega \right\}\\
\end{multline}
where, the angular integration measure in spherical polar coordinates $d\Omega = {\hbox{sin}}\theta d\theta d\phi$, as usual, and the radial integrals $A(l,m)$ and $B(l,m)$ are given by the equations,
\begin{equation}
   A(l,m)=i^{l} Y_{lm}(\hat{k})\int_{\lambda_0}^{\infty}  r^2 \frac{j_{l}(kr)}{r^3} dr
\end{equation}
and,
\begin{equation}
    B(l,m)=i^{l} Y_{lm}(\hat{k})\int_{\lambda_0}^{\infty}  r^2 \frac{j_{l}(kr)}{r^4} dr
\end{equation}
Here, $\lambda_0$ is the radius of the sphere bounding the ball $B^3$ which encloses the knot.
\subsubsection{The radial integrals}
In order to compute $A(l,m)$, we use the well-known relation between the spherical Bessel function and the Bessel function {\it viz.} 
\\
\begin{equation*}
j_{l}(kr)=\sqrt{\frac{\pi}{2kr}}J_{l+\frac{1}{2}}(kr)
\end{equation*}
It follows that\\ 
\begin{equation}
    \int_{\lambda_0}^{\infty}  r^2 \frac{j_{l}(Kr)}{r^3} dr=
    \sqrt{\frac{\pi}{2}}\int_{K\lambda_0}^{\infty}r^{\frac{-3}{2}}J_{l+\frac{1}{2}}(r)dr
\end{equation}
\\
We compute this integral for Re$ \lambda < \frac{1}{2} $ using the following 
formula \cite{Prudnikov2018},
\\
\begin{equation}
    \int_{a}^{\infty} x^{\lambda}J_{v}(x)dx
\end{equation}
\\
\begin{equation}
=2^{\lambda}\Gamma [\begin{matrix}
    (v+\lambda+1)/2\\
    (v-\lambda+1)/2
\end{matrix}]-\frac{a^{\lambda+v+1}}{2^v(\lambda+v+1)\Gamma(v+1)}{}_1 F_2(\frac{\lambda+v+1}{2};v+1,\frac{\lambda+v+3}{2};-\frac{a^2}{4})
\end{equation}
where, ~$\Gamma [\begin{matrix}
    (v+\lambda+1)/2\\
    (v-\lambda+1)/2
\end{matrix}]$=
$\frac{\Gamma (\frac{v+\lambda+1}{2})}{\Gamma( \frac{v-\lambda+1}{2})}$.
\\

In the above equation $\Gamma$ is the standard $\Gamma$-function and $F$ is the hypergeometric function.

In our case, $\lambda=-\frac{3}{2},v=l+\frac{1}{2},a=k \lambda_0.$
Therefore,
\begin{equation}
    A(l,m)=i^{l}\sqrt{\frac{\pi}{2}}Y_{l m}(\hat{k}
    )\left\{2^{-\frac{3}{2}} \Gamma [\begin{matrix}
    l/2\\
    (l+3)/2
\end{matrix}]-\frac{(k\lambda_0)^l}{ 2^{l+\frac{1}{2}}  l\Gamma(l+\frac{3}{2})}{}_1 F_2(\frac{l}{2};l+\frac{3}{2},\frac{l+2}{2};-\frac{k^2 \lambda_0^2}{4})\right\}
\end{equation}
In a similar way we get,
\begin{multline}
    B(l,m)=\\
    i^{l}K Y_{lm}(\hat{k}) \sqrt{\frac{\pi}{2}}\left\{2^{-\frac{5}{2}} \Gamma [\begin{matrix}
    (l-1)/2\\
    (l+4)/2
\end{matrix}]-\frac{(k\lambda_0)^{l-1}}{ 2^{l+\frac{1}{2}}  (l-1)\Gamma(l+\frac{3}{2})}{}_1 F_2(\frac{l-1}{2};l+\frac{3}{2},\frac{l+1}{2};-\frac{k^2 \lambda_0^2}{4})\right\}
\end{multline}
\subsubsection{The angular integrals}
The angular integrals contain terms which depend on products of upto three $R_{j}$'s where,
\begin{align}
R_1 = \alpha=\hbox{Sin}\theta \hbox{Cos}\phi=\sqrt{\frac{2\pi}{3}}(Y_{1,-1}-Y_{1,1}) \nonumber \\
R_2 = \beta=Sin\theta Sin\phi=i\sqrt{\frac{2\pi}{3}}(Y_{1,-1}+Y_{1,1}) \nonumber \\
R_3 =\gamma=Cos\theta=\sqrt{\frac{4\pi}{3}}Y_{1,0}
\end{align}
To solve for the angular integrals, we adopt the following strategy: 
Integrals involving one or
two spherical harmonics can be performed by using the orthonormality properties of the spherical harmonics.
\\
\begin{equation}
    \int_{\theta=0}^{\pi} \int_{\phi=0}^{2 \pi} Y^*_{l,m}(\theta,\phi) Y_{l',m'}(\theta,\phi)~d\Omega=\delta_{l,l'} \delta_{m,m'}
\end{equation}
Integrals involving three or more spherical harmonics can be solved by a repeated application of the addition theorem of spherical harmonics which enables us to express the product of two spherical harmonics as a sum over a single spherical harmonic multiplied by the appropriate Clebsch-Gordon coefficient. 
\begin{multline}
     Y_{l_1,m_1}Y_{l_2,m_2}\\
     =\sum_{L=|l_1-l_2|}^{(l_1+l_2)} \sum_{M=-L}^{+L} \sqrt{\frac{(2l_1+1)(2l_2+1)}{4\pi (2L+1)}}<l_1,l_2;0,0|L,0> <l_1,l_2;m_1,m_2|L,M> Y_{L,M}
\end{multline}
For example, the angular integral over the product of three spherical harmonics is given by
\begin{multline}
    \int_{\phi=0}^{2\pi} \int_{\theta=0}^{\pi} Y^{*}_{l, m}(\theta,\phi)Y_{l_{1}, m_{1}}(\theta,\phi)Y_{l_{2}, m_{2}}(\theta,\phi)~ d\Omega\\
    =\sqrt{\frac{(2l_{1}+1)(2l_{2}+1)}{4\pi(2l+1)}}<l_{1},l_{2};0,0|l,0><l_{1},l_{2};m_{1},m_{2}|l,m>
\end{multline}

\subsection{Calculation of $V_{ni}$}
\subsubsection{Terms involving an even number of $R_i$s:}
Angular integrals appearing in $V_{ni}$, equation \eqref{28}, with zero or two $R_i$s, (correspondingly, one or three spherical harmonics), can be 
obtained from six basic integrals, by using the 
following conditions satisfied by the Clebsch-Gordon coefficients:
\begin{equation}
m_{1}+m_{2}=m~~{\hbox{and}}~~|l_{1}-l_{2}|\le l \le l_{1}+l_{2}
\end{equation}
We now illustrate the computation of a generic  integral involving three 
spherical harmonics which will be useful in what follows.\\
Let's consider
\begin{multline}
\sum_{l=0}^{\infty}\sum_{m=-l}^{m=l} A(l,m) \int Y^{*}_{l,m}(\theta,\phi)Y_{1,0}(\theta,\phi)Y_{1, 0}(\theta,\phi)~ d\Omega\\ 
=\sum_{l=0}^{\infty}\sum_{m=-l}^{m=l} A(l,m)\sqrt{\frac{(2 \times 1+1)(2 \times 1+1)}{4\pi(2l+1)}}<1,1;0,0|l,0><1,1;0,0|l,m>
\end{multline}
Here, $l_{1}=l_{2}=1,m_{1}= m_{2}=0$. So the only possible vales for $l$ are $0\le l \le 2.$ {\it i.e,} $l=0,1,2$, and $m$ is always zero as can be seen from equation (44). Therefore, the  only terms which survive in the infinite sum of equation (45) are
\\

$=A(0,0)\sqrt{\frac{(2 \times 1+1)(2 \times 1+1)}{4\pi(2 \times 0+1)}}<1,1;0,0|0,0><1,1;0,0|0,0>\\+A(1,0)\sqrt{\frac{(2 \times 1+1)(2 \times 1+1)}{4\pi(2 \times 1+1)}}<1,1;0,0|1,0><1,1;0,0|1,0>\\+A(2,0)\sqrt{\frac{(2* \times 1+1)(2 \times 1+1)}{4\pi(2 \times 2+1)}}<1,1;0,0|2,0><1,1;0,0|2,0>$
\\
$=A(0,0)\frac{3}{\sqrt{4\pi}}<1,1;0,0|0,0><1,1;0,0|0,0>\\+A(1,0)\frac{3}{\sqrt{12\pi}}<1,1;0,0|1,0><1,1;0,0|1,0>\\+A(2,0)\frac{3}{\sqrt{20\pi}}<1,1;0,0|2,0><1,1;0,0|2,0>$
\\
Putting in the values of   Clebsch-Gordon coefficients we get,
\\
$=A(0,0)\frac{3}{\sqrt{4\pi}}(-\frac{1}{\sqrt{3}})^2+A(1,0)\frac{3}{\sqrt{12\pi}}\times 0+A(2,0)\frac{3}{\sqrt{20\pi}}(\sqrt{\frac{2}{3}})^2$
\\
$=A(0,0)\sqrt{\frac{1}{4\pi}}+A(2,0)\sqrt{\frac{1}{5\pi}}$
\\

In a  similar way, the following integrals can be computed:
\\
\begin{equation}
\sum_{l=0}^{\infty}\sum_{m=-l}^{m=l} A(l,m)\int Y^{*}_{l,m}(\theta,\phi)Y_{1,-1}(\theta,\phi)Y_{1, -1}(\theta,\phi)~d \Omega=A(2,-2)\sqrt{\frac{3}{10\pi}}
\end{equation}
\begin{equation}
\sum_{l=0}^{\infty}\sum_{m=-l}^{m=l} A(l,m)\int Y^{*}_{l,m}(\theta,\phi)Y_{1,1}(\theta,\phi)Y_{1, 1}(\theta,\phi)~d\Omega=A(2,2)\sqrt{\frac{3}{10\pi}}
\end{equation}
\begin{equation}
\sum_{l=0}^{\infty}\sum_{m=-l}^{m=l} A(l,m)\int Y^{*}_{l,m}(\theta,\phi)Y_{1,0}(\theta,\phi)Y_{1, 1}(\theta,\phi)~d\Omega=A(2,1)\sqrt{\frac{3}{20\pi}}
\end{equation}
\begin{equation}
\sum_{l=0}^{\infty}\sum_{m=-l}^{m=l} A(l,m)\int Y^{*}_{l,m}(\theta,\phi)Y_{1,0}(\theta,\phi)Y_{1,-1}(\theta,\phi)~d\Omega=A(2,-1)\sqrt{\frac{3}{20\pi}}
\end{equation}
\begin{multline}
\sum_{l=0}^{\infty}\sum_{m=-l}^{m=l} A(l,m)\int Y^{*}_{l,m}(\theta,\phi)Y_{1,1}(\theta,\phi)Y_{1,-1}(\theta,\phi)~d\Omega=\\
A(2,0)\frac{1}{\sqrt{20\pi}}-A(0,0)\frac{1}{\sqrt{4 \pi}}
\end{multline}
\\
Recalling that each $R_i$, and hence the product of two or more $R_i$s,
can be expressed in terms of spherical harmonics, the relevant integrals that appear in the computation of $V_{ni}$ can be read off
by comparing with the above basic integrals (Appendix A).They are given by the following equations:
\begin{multline}
\sum_{l=0}^{\infty}\sum_{m=-l}^{m=l} A(l,m) \int Y^{*}_{l m}(\theta,\phi) R_1 R_1~ d\Omega \\
=A(2,-2)\sqrt{\frac{2\pi}{15}}+A(2,2)\sqrt{\frac{2\pi}{15}}-A(2,0)\frac{2}{3}\sqrt{\frac{\pi}{5}}+\frac{2 \sqrt{\pi}}{3}A(0,0)
\end{multline}
\begin{multline}
\sum_{l=0}^{\infty}\sum_{m=-l}^{m=l} A(l,m) \int Y^{*}_{l m}(\theta,\phi) R_2 R_2~ d\Omega  \\
=-A(2,-2)\sqrt{\frac{2\pi}{15}}-A(2,2)\sqrt{\frac{2\pi}{15}}-A(2,0)\frac{2}{3}\sqrt{\frac{\pi}{5}}+\frac{2 \sqrt{\pi}}{3}A(0,0)
\end{multline}
\begin{equation}
      \sum_{l=0}^{\infty}\sum_{m=-l}^{m=l} A(l,m) \int Y^{*}_{l m}(\theta,\phi) R_3 R_3 ~ d\Omega
      =A(0,0)\frac{2}{3}\sqrt{\pi}+A(2,0)\frac{4}{3}\sqrt{\frac{\pi}{5}}
\end{equation}
\begin{equation}
\sum_{l=0}^{\infty}\sum_{m=-l}^{m=l} A(l,m) \int Y^{*}_{l m}(\theta,\phi)R_1 R_2~ d\Omega 
=i\frac{2\pi}{3}\left( A(2,-2)\sqrt{\frac{3}{10\pi}}-A(2,2)\sqrt{\frac{3}{10\pi}}\right)
\end{equation}
\begin{equation}
\sum_{l=0}^{\infty}\sum_{m=-l}^{m=l} A(l,m) \int Y^{*}_{l m}(\theta,\phi) R_1 R_3 d\Omega 
=A(2,-1)\sqrt{\frac{2\pi}{15}}-A(2,1)\sqrt{\frac{2\pi}{15}}
\end{equation}
\begin{equation}
\sum_{l=0}^{\infty}\sum_{m=-l}^{m=l} A(l,m) \int Y^{*}_{l m}(\theta,\phi) R_2 R_3~ d\Omega  
=i A(2,-1)\sqrt{\frac{2\pi}{15}}+ i A(2,1)\sqrt{\frac{2\pi}{15}}
\end{equation}
\begin{align}
\sum_{l=0}^{\infty}\sum_{m=-l}^{m=l} A(l,m) \int Y^{*}_{l m}(\theta,\phi) d\Omega
    &=\sum_{l=0}^{\infty}\sum_{m=-l}^{m=l} A(l,m) \times \sqrt{4 \pi} \delta_{l,0} \delta_{m,0} \nonumber \\
    &=\sqrt{4 \pi}A(0,0)
\end{align}
\\
\subsubsection{Terms involving an odd number of $R_i$s:}
Angular integrals appearing in $V_{ni}$ with one or three $R_i$s (and hence two or four spherical harmonics), can be handled similarly. The one with two 
spherical harmonics can be calculated through the orthonormality condition.
\begin{equation}
\int \sum_{l,m}B(l,m) Y^{*}_{l,m}(\theta,\phi)R_1~d\Omega=\sqrt{\frac{2\pi}{3}}[B(1,-1)-B(1,1)]
\end{equation}
\begin{equation}
\int \sum_{l,m}B(l,m) Y^{*}_{l,m}(\theta,\phi)R_2~d\Omega=i\sqrt{\frac{2\pi}{3}}[B(1,-1)+B(1,1)]
\end{equation}
\begin{equation}
\int \sum_{l,m}B(l,m) Y^{*}_{l,m}(\theta,\phi)R_3~d\Omega=\sqrt{\frac{4\pi}{3}}B(1,0)
\end{equation}
\\
As already mentioned, the integrals involving three $R_i$s and hence four spherical harmonics, can be simplified by using the addition theorem of spherical harmonics, followed by the orthonormality condition.  
\\
Substituting the simplifications of the spherical harmonic products from Appendix B and performing the integrals yields, the following contributions to $V_{ni}$,
\begin{multline}\int \sum_{l,m}B(l,m) Y^{*}_{l,m}(\theta,\phi)R_1 R_1 R_1 ~d\Omega\\
=(\frac{2\pi}{3})^\frac{3}{2}[\frac{3}{2 \pi}\sqrt{\frac{3}{70}}~B(3,-3)+\frac{9}{10 \pi}B(1,-1)-\frac{9}{10\pi}\sqrt{\frac{1}{14}}B(3,-1)\\-\frac{9}{10 \pi}
~B(1,1)+\frac{9}{10 \pi \sqrt{14}}~B(3,1)-\frac{3}{2 \pi}\sqrt{\frac{3}{70}}  B(3,3)]
\end{multline}
\begin{multline}
\int \sum_{l,m}B(l,m) Y^{*}_{l,m}(\theta,\phi)R_2 R_2 R_2 ~d\Omega\\=
-i(\frac{2\pi}{3})^\frac{3}{2}[\frac{3}{2 \pi}\sqrt{\frac{3}{70}}~B(3,-3)-\frac{9}{10 \pi}B(1,-1)+\frac{9}{10\pi}\sqrt{\frac{1}{14}}~B(3,-1)\\
-\frac{9}{10 \pi}~B(1,1)+\frac{9}{10 \pi \sqrt{14}}~B(3,1)+\frac{3}{2 \pi}\sqrt{\frac{3}{70}} ~B(3,3)]
\end{multline}
\begin{multline}
\int \sum_{l,m}B(l,m) Y^{*}_{l,m}(\theta,\phi)R_3 R_3 R_3 ~d\Omega\\
=(\frac{4 \pi}{3})^\frac{3}{2}\left\{\frac{9}{20 \pi}B(1,0)+\frac{3}{10 \pi}\sqrt{\frac{3}{7}}B(3,0)\right\}
\end{multline}
\begin{multline}
\int \sum_{l,m}B(l,m) Y^{*}_{l,m}(\theta,\phi)R_1 R_2 R_2 ~d\Omega\\
=-(\frac{2\pi}{3})^\frac{3}{2}[\frac{3}{2 \pi}\sqrt{\frac{3}{70}}~B(3,-3)-\frac{3}{10 \pi}B(1,-1)+\frac{3}{10\pi}\sqrt{\frac{1}{14}}B(3,-1)\\
+\frac{3}{10 \pi}B(1,1)-\frac{3}{10 \pi \sqrt{14}}B(3,1)-\frac{3}{2 \pi}\sqrt{\frac{3}{70}} B(3,3)]
\end{multline}
\begin{multline}
\int \sum_{l,m}B(l,m) Y^{*}_{l,m}(\theta,\phi)R_1 R_1 R_2 ~d\Omega\\
=i (\frac{2 \pi}{3})^\frac{3}{2} [\frac{3}{2 \pi}\sqrt{\frac{3}{70}}~B(3,-3)+\frac{3}{10 \pi}B(1,-1)-\frac{3}{10\pi}\sqrt{\frac{1}{14}}B(3,-1)\\
+\frac{3}{10 \pi}B(1,1)-\frac{3}{10 \pi \sqrt{14}}B(3,1)+\frac{3}{2 \pi}\sqrt{\frac{3}{70}} B(3,3)]
\end{multline}
\begin{multline}
\int \sum_{l,m}B(l,m) Y^{*}_{l,m}(\theta,\phi)R_1 R_3 R_3 ~d\Omega\\
=\frac{1}{5}\sqrt{\frac{2\pi}{3}}B(1,-1)+\frac{4}{5}\sqrt{\frac{\pi}{21}}B(3,-1)-\frac{1}{5}\sqrt{\frac{2\pi}{3}}B(1,1)-\frac{4}{5}\sqrt{\frac{\pi}{21}} B(3,1)
\end{multline}
\begin{multline}
\int \sum_{l,m}B(l,m) Y^{*}_{l,m}(\theta,\phi)R_1 R_1 R_3 ~d\Omega\\=\sqrt{\frac{2 \pi}{105}} B(3,-2)+\frac{2}{5}\sqrt{\frac{\pi}{3}}B(1,0)-\frac{2}{5}\sqrt{\frac{\pi}{7}}B(3,0)+\sqrt{\frac{2 \pi}{105}}B(3,2)
\end{multline}
\begin{multline}
\int \sum_{l,m}B(l,m) Y^{*}_{l,m}(\theta,\phi)R_2 R_2 R_3 ~d\Omega\\=-\sqrt{\frac{2 \pi}{105}} B(3,-2)+\frac{2}{5}\sqrt{\frac{\pi}{3}}B(1,0)
-\frac{2}{5}\sqrt{\frac{\pi}{7}}B(3,0)-\sqrt{\frac{2 \pi}{105}}B(3,2)
\end{multline}
\begin{multline}
\int \sum_{l,m}B(l,m) Y^{*}_{l,m}(\theta,\phi)R_2 R_3 R_3 ~d\Omega\\
=\frac{i}{5}\sqrt{\frac{2 \pi}{3}}B(1,-1)+\frac{4i}{5}\sqrt{\frac{\pi}{21}}B(3,-1)\\+\frac{i}{5}\sqrt{\frac{2 \pi}{3}}B(1,1)+\frac{4i}{5}\sqrt{\frac{\pi}{21}}B(3,1)
\end{multline}
\begin{equation}
\int \sum_{l,m}B(l,m) Y^{*}_{l,m}(\theta,\phi)R_1 R_2 R_3 ~d\Omega 
=i\sqrt{\frac{2 \pi}{105}}B(3,-2)-i\sqrt{\frac{2 \pi}{105}}B(3,2)
\end{equation}
Gathering all the integrals calculated above, we may reorganise the expression for $V_{ni}$ as follows:
\begin{equation}
\label{96}
    V_{ni}=\sum_{j=1}^{4} V_{ni}^{(j)}
\end{equation}
A few comments are in order. First, the $V_{ni}^{(j)}(\hat{k}_i,\hat{k}_n),j=1,2,3,4$ are functions of the unit vector of $(\Vec{k}_i-\Vec{k}_n)$. Second, $V_{ni}^{(1)}$ and $V_{ni}^{(2)}$ are contributions to the $S$-matrix coming from the quadrupole moment, the latter coming from the trace part. Similarly, the remaining two are contributions to the $S$-matrix coming from the octopole moment. Third, the explicit expressions for $V_{ni}^{(j)}$ are tedious and not very illuminating, and hence relegated to Appendix C. It should be noted from those expressions, however that, the coefficients containing $A(l,m)$ and $B(l,m)$ are identical for all knots. Only the multipole moments.e.g,~$Q^{r_i}_{R_j R_k},Q^{r_i},O^{r_i}_{R_jR_k R_l},O^{r_i}_{R_j}$ are different for different knots. Finally the S-matrix can be expressed in terms of the $V_{ni}^{(j)}$ by simple substitution, and takes the form
 \begin{multline}
        S_{ni}(\hat{k}_i,\hat{k}_n)\\
        =\delta_{ni}-2 \pi i \delta (E_n-E_i)\left[V_{ni}^{(1)}(\hat{k}_i,\hat{k}_n)+V_{ni}^{(2)}(\hat{k}_i,\hat{k}_n)+V_{ni}^{(3)}(\hat{k}_i,\hat{k}_n)+V_{ni}^{(4)}(\hat{k}_i,\hat{k}_n)\right]
    \end{multline}
The explicit display of the contribution of different multipoles to the
S-matrix will turn out to be useful when we specialize to the case of a torus knot, which we discuss in the next section. 
\section{S-Matrix For Torus Knots And Its Factorization}\label{4}
In this section, we will use the general formalism developed in the previous section to calculate the 
$S$-matrices for the scattering of a charged particle from a particular class of knots -- the so-called $(p,q)$ torus knots.
\\ 

A $(p,q)$ torus knot is defined by a closed loop on a putative torus which winds $p$ and $q$ times around the two inequivalent cycles of the torus respectively, with $p$ and $q$ being relatively prime. It
may be parametrized as follows:\\
\begin{equation}
\label{76}
    x'(t)=(2+ {\hbox{cos}}qt){\hbox{cos}}pt; \:\:
    y'(t)=(2+{\hbox{cos}}qt){\hbox{sin}}pt; \:\:
    z'(t)=-{\hbox{sin}qt}
\end{equation}
\\
Here, $t$ is a  parameter whose domain is $0 \le t \le 2 \pi$, and the knot lies on the surface of the torus given by the equation,
$(r'-2)^2+z'^2=1$ in cylindrical coordinates. As usual $(r')^2 = (x')^2 +(y')^2$. 

\subsection {Multipole Moments for $(p,q)$ Torus Knots} 

\subsubsection{Quadrupole Moments} As already mentioned in \eqref{10} to \eqref{13}, it suffices to calculate the three 
numbers $K^1,K^2, K^3$ to find all the quadrupole moments . Using the above parametrization, for a $(p,q)$ knot, these turn out to be 
\begin{equation}
    K^1 =\int _{0}^{2 \pi} z' \frac{dy'}{dt}dt =0
\end{equation}
\begin{equation}
K^2 =\int _{0}^{2 \pi} x' \frac{dz'}{dt}dt =0
\end{equation}
\begin{equation}
K^3 =\int _{0}^{2 \pi} y' \frac{dx'}{dt}dt =-\frac{9 p \pi}{2}
\end{equation}
\\
It follows that the non-vanishing quadrupole moments obtained from \eqref{8} to \eqref{9}, are given by
\\

\begin{equation}
\label{73}
Q^{x}_{\alpha\gamma}=Q^{y}_{\beta\gamma}=-Q^{z}_{\alpha\alpha}=-Q^{z}_{\beta\beta}=\frac{9p \pi}{2}~~~\hbox{and}~~
~Q^{z}=-\frac{18p \pi }{2}
\end{equation}
The general expressions for the quadrupole moments of a $(p,q)$ knot may be used to read off the corresponding expressions for an unknot. 

Let us consider an unknot, $unknot_1$, of radius $r=3$ in the $x-y$ plane. The parametric equation for this unknot is,
\\
\begin{equation}
\label{74}
 x'(t)=3 \hbox{cos}t, ~~ y'(t)=3 \hbox{sin}t, ~~ z'(t)=0
 \end{equation} 
 \\
The corresponding values for $K^i$ turn out to be ~$K^1=0, K^2=0 $ and $K^3=-9 \pi$.
 \\
The non-vanishing quadrupole moments for this unknot are given by,
 \\
 \begin{equation}
 \label{75}
Q^{x}_{\alpha\gamma}=Q^{y}_{\beta\gamma}=-Q^{z}_{\alpha\alpha}=-Q^{z}_{\beta\beta}=9 \pi~~~\hbox{and}~~
~Q^{z}=-18 \pi
\end{equation}
Comparing equations \eqref{73} and \eqref{75}, we get the following curious relations between the quadrupole moments for a $(p,q)$ torus knot and an unknot:
 \begin{equation}
 \label{76}
     Q^{r_i}_{R_j R_k}(p,q)=\frac{p}{2} \times Q^{r_i}_{R_j R_k} (unknot_1)
 \end{equation}
 \begin{equation}
 \label{77}
     Q^{r_i}(p,q)=\frac{p}{2} \times Q^{r_i}(unknot_1)
 \end{equation}
 \subsubsection{Octopole Moments}
The non-vanishing octopole moments for a $(p,q)$ torus knot can be obtained from equation \eqref{16} and \eqref{17} by straightforward integration and are given below.
\\
For the x-component,
\begin{equation}
\int_K O^{x}_{\beta^3}dt=\int_K O^{x}_{\alpha^2 \beta}dt=\int_K O^{x}_{\beta \gamma^2}dt=2 q \pi ~~\hbox{and}~~\int_K O^{x}_{\beta}dt=10 q \pi
\end{equation}
For the y-component,
\begin{equation}
\int_k O^{y}_{\alpha^3}dt=\int_K O^{y}_{\alpha \beta^2} dt=\int_K O^{y}_{\alpha \gamma^2}dt=-2q \pi~~\hbox{and}~~
\int_K O^{y}_{\alpha}dt=-10 q \pi
\end{equation}
For the z-component,
\begin{equation}
\int_K O^{z}_{\alpha \beta \gamma}dt=0
\end{equation}
Now consider a circular unknot, $unknot_2$, of unit radius in the $xz$ plane with parametric equation
\begin{equation}
\label{88}
x'(t)=2+\hbox{cos}t,~~ y'(t)=0,~~z'(t)=\hbox{sin}t
\end{equation}
The non-vanishing octopole moments for this unknot are, for the 
$x$- component,
\begin{equation}
    \int_K O^{x}_{\alpha^2 \beta} dt=-4 \pi~~\hbox{and}~~\int_K O^{x}_{\beta} dt=-4 \pi
\end{equation}
for the $y$-component
\begin{equation}
    \int_K O^{y}_{\alpha^3}dt=\int_K O^{y}_{\alpha \gamma^2}dt=4 \pi ~\hbox{and} \int_K O^{y}_{\alpha} dt=16 \pi
\end{equation}
and for the $z$-component,
\begin{equation}
    \int_K O^{z}_{\alpha \beta \gamma}  dt=-4 \pi
\end{equation}
Now, consider a third  unknot, $unknot_3$, of unit radius in $y-z$ 
plane by setting 
\begin{equation}
\label{92}
    x'(t)=0, ~~ y'(t)=(2+cost),~~z'(t)=sint
\end{equation}
The non-vanishing octopole moments for this unknot are,
\\
For the x-component,
\begin{equation}
    \int_{K} O^{x}_{\beta^3}dt=\int_{K} O^{x}_{\beta \gamma^2} dt=-4 \pi~~\hbox{and}~~\int_{K} O^{x}_{\beta} dt=-16\pi
\end{equation}
for the y-component,
\begin{equation}
    \int_{K} O^{y}_{\alpha \beta^2} dt=\int_{k} O^{y}_{\alpha} dt=4 \pi
\end{equation}
and for the z-component,
\begin{equation}
    \int_{K} O^{z}_{\alpha \beta \gamma} dt=4 \pi
\end{equation}
It is evident that the 
non-vanishing octopole moments for a $(p,q)$ knot can be written as a linear combination of the corresponding moments for the two specified unknots
\begin{equation}
\label{89}
    \int_{K}O^{r_i}_{R_jR_k R_l}(p,q)dt=-\frac{q}{2} \times \left\{\int_{k}O^{r_i}_{R_jR_k R_l}(unknot_2)dt+\int_{K}O^{r_i}_{R_jR_k R_l}(unknot_3)dt\right\}
\end{equation}
\begin{equation}
\label{90}
    \int_{K}O^{r_i}_{R_j}(p,q)dt=-\frac{q}{2} \times \left\{\int_{k}O^{r_i}_{R_j}(unknot_2)dt+\int_{K}O^{r_i}_{R_j}(unknot_3)dt\right\}
\end{equation}
\subsection{S-matrix And Its Factorization}
The decomposition of the multipole moments of a torus knot into the multipole moments of unknots delineated above carries over to the S-matrix:
\begin{multline}
\label{125}
        S_{ni}(p,q)\\
        =\delta_{ni}-2 \pi i \delta (E_n-E_i)\left[V_{ni}^{(1)}(p,q)+V_{ni}^{(2)}(p,q)+V_{ni}^{(3)}(p,q)+V_{ni}^{(4)}(p,q)\right]
    \end{multline}
which can be re-written in the following form, by putting \eqref{76},\eqref{77},\eqref{89},\eqref{90},
\begin{multline}
\label{92}
        S_{ni}(p,q) \\
        =\delta_{ni}-2 \pi i \delta (E_n-E_i)[\frac{p}{2} \times \left\{V_{ni}^{(1)}(unknot_1)+V_{ni}^{(2)}(unknot_1)\right\}\\
        -\frac{q}{2}\left\{V_{ni}^{(3)}(unknot_2)+V_{ni}^{(4)}(unknot_2)\right\} \\
        -\frac{q}{2}\left\{V_{ni}^{(3)}(unknot_3)+V_{ni}^{(4)}(unknot_3)\right\}]
\end{multline}
The linear relationship between the S-matrix and the matrix $V_{ni}$ in the first order Born approximation, the universality of the radial integrals, and the decomposition of the multipole moments of $(p,q)$ torus knots into multipole moments of a triad of unknots, allow us to arrive at the above factorization property. It is curious to note that it is reminiscent of the skein relations satisfied by knot diagrams \cite{Livingston1993}\cite{Atiyah1990}.\\ 
\section{Conclusions and Outlook}
In this paper we have studied the Aharonov-Bohm scattering of a charged particle from a knotted solenoid. The vector potential due to the knotted solenoid is computed in a multipole expansion, and used to further compute the S-matrix in the Born approximation. It is found that to capture the signature of the knottedness of the solenoid, we need to go beyond the quadrupole moment, upto the octopole moment. After setting up the general formalism, it is applied to the specific case of a $(p,q)$ torus knot. A curious factorization property of the multipole moments allowed us to express the S-matrix of the torus knot as a combination of the S-matrices of a triad of unknots, one each in the $xy$, $yz$, and $zx$ planes.\\ 

The general formalism developed in this paper relies on the multipole expansion of the vector potential, a trick necessitated by the absence of any symmetry that can be exploited, as in the standard Aharonov-Bohm effect. It should be interesting to use this trick to study Aharonov-Bohm scattering from linked solenoids for example. It will also be interesting to study these scattering processes in quantum field theory using standard techniques, in particular to study scattering of fermions from knotted and linked solenoids. We hope to return to these issues in the near future.
\section{Acknowledgement} This work is partially funded by a grant from the Infosys Foundation. \\

\section{Appendix}
\subsection{Appendix A}
Here, we enumerate all possible products of $R_j$'s that appear in the angular integrals in the expression of $V_{ni}$, and write them in terms of spherical harmonics, which will allow us to compute the $V_{ni}$ and hence the $S$-matrix explicitly.\\ 
\begin{equation}
R_1 R_1=\alpha^2=\frac{2\pi}{3}(Y_{1,-1}Y_{1,-1}+Y_{1,1}Y_{1,1}-2Y_{1,1}Y_{1,-1})
\end{equation}
\begin{equation}
    R_2 R_2=\beta^2=-\frac{2\pi}{3}(Y_{1,-1}Y_{1,-1}+Y_{1,1}Y_{1,1}+2Y_{1,1}Y_{1,-1})
    \end{equation}
    \begin{equation}
        R_3 R_3=\gamma^2=\frac{4\pi}{3}Y_{1,0}Y_{1,0}
    \end{equation}
    \begin{equation}
    R_1 R_2=\alpha \beta=i\frac{2\pi}{3}(Y_{1,-1}Y_{1,-1}-Y_{1,1}Y_{1,1})
    \end{equation}
    \begin{equation}
    R_1 R_3=\alpha \gamma=\frac{2\sqrt{2}\pi}{3}(Y_{1,-1}Y_{1,0}-Y_{1,1}Y_{1,0})
    \end{equation}
    \begin{equation}
    R_2 R_3=\beta \gamma=i\frac{2\sqrt{2}\pi}{3}(Y_{1,-1}Y_{1,0}+Y_{1,1}Y_{1,0})
    \end{equation}
    \begin{equation}
    R_1 R_1 R_1=\alpha^3=(\frac{2\pi}{3})^\frac{3}{2}(Y_{1,-1}^3-3Y_{1,-1}^2Y_{1,1}+3Y_{1,-1}Y_{1,1}^2-Y_{1,1}^3)
    \end{equation}
    \begin{equation}
    R_2 R_2 R_2=\beta^3=-i(\frac{2\pi}{3})^\frac{3}{2}(Y_{1,-1}^3+3Y_{1,-1}^2Y_{1,1}+3Y_{1,-1}Y_{1,1}^2+Y_{1,1}^3)
    \end{equation}
    \begin{equation}
    R_3 R_3 R_3=\gamma^3=(\frac{4 \pi}{3})^\frac{3}{2} Y_{1,0}Y_{1,0}Y_{1,0}
    \end{equation}
    \begin{multline}
    R_1 R_2 R_2=\alpha \beta^2=-(\frac{2\pi}{3})^\frac{3}{2}(Y_{1,-1}Y_{1,-1}Y_{1,-1}
    +Y_{1,-1}Y_{1,-1}Y_{1,1}\\-Y_{1,1}Y_{1,1}Y_{1,-1}-Y_{1,1}Y_{1,1}Y_{1,1})
    \end{multline}
    \begin{multline}
    R_1 R_1 R_2=\alpha^2 \beta=i (\frac{2 \pi}{3})^\frac{3}{2} (Y_{1,-1}Y_{1,-1}Y_{1,-1}-Y_{1,1}Y_{1,-1}Y_{1,-1}\\-Y_{1,1}Y_{1,1}Y_{1,-1}+Y_{1,1}Y_{1,1}Y_{1,1})
    \end{multline}
    \begin{equation}
    R_1 R_3 R_3=\alpha \gamma^2=\frac{4 \pi}{3}\sqrt{\frac{2\pi}{3}}(Y_{1,-1}Y_{1,0}Y_{1,0}-Y_{1,1}Y_{1,0}Y_{1,0})
    \end{equation}
    \begin{equation}
    R_1 R_1 R_3=\alpha^2 \gamma=\frac{2 \pi}{3}\sqrt{\frac{4\pi}{3}}(Y_{1,-1}Y_{1,-1}Y_{1,0}-2Y_{1,1}Y_{1,-1}Y_{1,0}+Y_{1,1}Y_{1,1}Y_{1,0})
    \end{equation}
    \begin{equation}
    R_2 R_2 R_3=\beta^2 \gamma=-\frac{2 \pi}{3}\sqrt{\frac{4\pi}{3}}(Y_{1,-1}Y_{1,-1}Y_{1,0}+2Y_{1,1}Y_{1,-1}Y_{1,0}+Y_{1,1}Y_{1,1}Y_{1,0})
    \end{equation}
    \begin{equation}
    R_2 R_3 R_3=\beta \gamma^2=i\frac{4\pi}{3} \sqrt{\frac{2\pi}{3}}(Y_{1,-1}Y_{1,0}Y_{1,0}+Y_{1,1}Y_{1,0}Y_{1,0})
    \end{equation}
    \begin{equation}
    R_1 R_2 R_3=\alpha \beta \gamma=i\frac{2 \pi}{3}\sqrt{\frac{4 \pi}{3}}(Y_{1,-1}Y_{1,-1}Y_{1,0}-Y_{1,1}Y_{1,1}Y_{1,0})
    \end{equation}
\subsection{Appendix B}
Here, we enumerate the combinations of three spherical harmonics which contribute to $V_{ni}$,especially integrals involving products of three $R_j$'s,\\
\begin{equation}
Y_{1,0}Y_{1,1}Y_{1,1}=\frac{3}{2 \pi \sqrt{70}} Y_{3,2}
\end{equation}
\begin{equation}
Y_{1,1}Y_{1,1}Y_{1,1}=\frac{3}{2 \pi}\sqrt{\frac{3}{70}} \times Y_{3,3}
\end{equation}
\begin{equation}
    Y_{1,-1}Y_{1,1}Y_{1,1}=-\frac{3}{10 \pi}Y_{1,1}+\frac{3}{10 \pi \sqrt{14}}Y_{3,1}
\end{equation}
\begin{equation}
    Y_{1,-1}Y_{1,-1}Y_{1,-1}=\frac{3}{2 \pi}\sqrt{\frac{3}{70}}~Y_{3,-3}
\end{equation}
\begin{equation}
    Y_{1,1}Y_{1,-1}Y_{1,-1}=-\frac{3}{10 \pi}Y_{1,-1}+\frac{3}{10\pi}\sqrt{\frac{1}{14}}~Y_{3,-1}
\end{equation}
\begin{equation}
    Y_{1,0}Y_{1,-1}Y_{1,-1}=\frac{3}{2 \pi}\sqrt{\frac{1}{70}}~Y_{3,-2}
\end{equation}
\begin{equation}
    Y_{1,-1}Y_{1,0}Y_{1,0}=\frac{3}{20 \pi}~Y_{1,-1}+\frac{3}{10\pi}\sqrt{\frac{2}{7}}~Y_{3,-1}
\end{equation}
\begin{equation}
    Y_{1,1}Y_{1,0}Y_{1,0}=\frac{3}{20 \pi}~Y_{1,1}+\frac{3}{10 \pi}\sqrt{\frac{2}{7}}~Y_{3,1}
\end{equation}
\begin{equation}
    Y_{1,0}Y_{1,0}Y_{1,0}=\frac{9}{20 \pi}~Y_{1,0}+\frac{3}{10 \pi}\sqrt{\frac{3}{7}}~Y_{3,0}
\end{equation}
\begin{equation}
    Y_{1,0}Y_{1,-1}Y_{1,1}=-\frac{3}{20 \pi}~Y_{1,0}+\frac{3}{20 \pi}\sqrt{\frac{3}{7}}~Y_{3,0}
\end{equation}
\subsection{Appendix C}
\subsubsection{Quadrupole Contribution to the S-matrix}

\begin{multline}
    V_{ni}^{(1)}=\\
    -\frac{\mu_0 \mathcal{M} e}{2MC}\sum_{j \le k=1}^{3}\left\{[\sum_{r_i=x,y,z}3 K_{r_i} Q^{r_i}_{R_j R_k}] \int \sum_{l,m}A(l,m) Y_{l,m}(\theta,\phi)R_j R_k d\Omega \right\}\\
    =-\frac{\mu_0 \mathcal{M} e}{2MC}\left[\sum_{r_i=x,y,z} 3 K_{r_i} Q^{r_i}_{R_1 R_1}\right] \times \left\{A(2,-2)\sqrt{\frac{2\pi}{15}}+A(2,2)\sqrt{\frac{2\pi}{15}}-A(2,0)\frac{2}{3}\sqrt{\frac{\pi}{5}}+\frac{2 \sqrt{\pi}}{3}A(0,0)\right\}\\
    -\frac{\mu_0 \mathcal{M} e}{2MC}\left[\sum_{r_i=x,y,z} 3 K_{r_i} Q^{r_i}_{R_2 R_2}\right] \times \left\{-A(2,-2)\sqrt{\frac{2\pi}{15}}-A(2,2)\sqrt{\frac{2\pi}{15}}-A(2,0)\frac{2}{3}\sqrt{\frac{\pi}{5}}+\frac{2 \sqrt{\pi}}{3}A(0,0)\right\}\\
    -\frac{\mu_0 \mathcal{M} e}{2MC}\left[\sum_{r_i=x,y,z} 3 K_{r_i} Q^{r_i}_{R_3 R_3}\right] \times \left\{A(0,0)\frac{2}{3}\sqrt{\pi}+A(2,0)\frac{4}{3}\sqrt{\frac{\pi}{5}}\right\}\\
     -\frac{\mu_0 \mathcal{M} e}{2MC}\left[\sum_{r_i=x,y,z} 3 K_{r_i} Q^{r_i}_{R_1 R_2}\right] \times \left\{i\frac{2\pi}{3}\times A(2,-2)\sqrt{\frac{3}{10\pi}}-i\frac{2\pi}{3}\times A(2,2)\sqrt{\frac{3}{10\pi}}\right\}\\
    -\frac{\mu_0 \mathcal{M} e}{2MC}\left[\sum_{r_i=x,y,z} 3 K_{r_i} Q^{r_i}_{R_1 R_3}\right] \times \left\{A(2,-1)\sqrt{\frac{2\pi}{15}}-A(2,1)\sqrt{\frac{2\pi}{15}}\right\}\\
    -\frac{\mu_0 \mathcal{M} e}{2MC}\left[\sum_{r_i=x,y,z} 3 K_{r_i} Q^{r_i}_{R_2 R_3}\right] \times \left\{i A(2,-1)\sqrt{\frac{2\pi}{15}}+ i A(2,1)\sqrt{\frac{2\pi}{15}}\right\}
\end{multline}
    \begin{multline}
        V_{ni}^{(2)}=-\frac{\mu_0 \mathcal{M} e}{2MC}\left[\sum_{r_i=x,y,z} K_{r_i}Q^{r_i}\right] \left\{\sum_{l,m} A(l,m) \int Y^{*}_{lm}(\theta,\phi) d\Omega \right\}\\
        =-\frac{\mu_0 \mathcal{M} e}{2MC} \left[\sum_{r_i=x,y,z} K_{r_i}Q^{r_i}\right] \times \sqrt{4 \pi}A(0,0)
    \end{multline}
    \subsubsection{Octopole Contribution to the S-matrix}
    \begin{multline}
        V_{ni}^{(3)}=\\
        -\frac{15}{2}\frac{\mu_0 \mathcal{M} e}{2MC}\sum_{j \le k \le l=1}^{3}\left\{[\sum_{r_i=x,y,z}K_{r_i} \int_{K} d\tau O^{r_i}_{R_j R_k R_l}] \int \sum_{l,m}B(l,m) Y^{*}_{l,m}(\theta,\phi)R_j R_k R_l d\Omega \right\}\\
        =-\frac{15}{2}\frac{\mu_0 \mathcal{M} e}{2MC}\left[\sum_{r_i=x,y,z} K_{r_i} \int_{K} d\tau O^{r_i}_{R_1 R_1 R_1}\right] \times (\frac{2\pi}{3})^\frac{3}{2}[\frac{3}{2 \pi}\sqrt{\frac{3}{70}}B(3,-3)+\frac{9}{10 \pi}B(1,-1)\\-\frac{9}{10\pi}\sqrt{\frac{1}{14}}B(3,-1)-\frac{9}{10 \pi}B(1,1)+\frac{9}{10 \pi \sqrt{14}}B(3,1)-\frac{3}{2 \pi}\sqrt{\frac{3}{70}}  B(3,3)]\\
        -\frac{15}{2}\frac{\mu_0 \mathcal{M} e}{2MC}\left[\sum_{r_i=x,y,z} K_{r_i} \int_{K} d\tau O^{r_i}_{R_2 R_2 R_2}\right] \times -i(\frac{2\pi}{3})^\frac{3}{2}[\frac{3}{2 \pi}\sqrt{\frac{3}{70}}B(3,-3)-\frac{9}{10 \pi}B(1,-1)\\
        +\frac{9}{10\pi}\sqrt{\frac{1}{14}}B(3,-1)-\frac{9}{10 \pi}B(1,1)+\frac{9}{10 \pi \sqrt{14}}B(3,1)+\frac{3}{2 \pi}\sqrt{\frac{3}{70}}  B(3,3)]\\
         -\frac{15}{2}\frac{\mu_0 \mathcal{M} e}{2MC}\left[\sum_{r_i=x,y,z} K_{r_i} \int_{K} d\tau O^{r_i}_{R_3 R_3 R_3}\right] \times \left((\frac{4 \pi}{3})^\frac{3}{2}\frac{9}{20 \pi}B(1,0)+(\frac{4 \pi}{3})^\frac{3}{2}\frac{3}{10 \pi}\sqrt{\frac{3}{7}}B(3,0)\right)\\
          -\frac{15}{2}\frac{\mu_0 \mathcal{M} e}{2MC}\left[\sum_{r_i=x,y,z} K_{r_i} \int_{K} d\tau O^{r_i}_{R_1 R_2 R_2}\right] \times-(\frac{2\pi}{3})^\frac{3}{2}[\frac{3}{2 \pi}\sqrt{\frac{3}{70}}B(3,-3)-\frac{3}{10 \pi}B(1,-1)\\
          +\frac{3}{10\pi}\sqrt{\frac{1}{14}}B(3,-1)+\frac{3}{10 \pi}B(1,1)-\frac{3}{10 \pi \sqrt{14}}B(3,1)-\frac{3}{2 \pi}\sqrt{\frac{3}{70}} B(3,3)]\\
          -\frac{15}{2}\frac{\mu_0 \mathcal{M} e}{2MC}\left[\sum_{r_i=x,y,z} K_{r_i} \int_{K} d\tau O^{r_i}_{R_1 R_1 R_2}\right] \times i (\frac{2 \pi}{3})^\frac{3}{2} [\frac{3}{2 \pi}\sqrt{\frac{3}{70}} B(3,-3)+\frac{3}{10 \pi}B(1,-1)\\
          -\frac{3}{10\pi}\sqrt{\frac{1}{14}}B(3,-1)+\frac{3}{10 \pi}B(1,1)-\frac{3}{10 \pi \sqrt{14}}B(3,1)+\frac{3}{2 \pi}\sqrt{\frac{3}{70}} B(3,3)]\\
          -\frac{15}{2}\frac{\mu_0 \mathcal{M} e}{2MC}\left[\sum_{r_i=x,y,z} K_{r_i} \int_{K} d\tau O^{r_i}_{R_1 R_3 R_3}\right] \times  [\frac{1}{5}\sqrt{\frac{2\pi}{3}}B(1,-1)+\frac{4}{5}\sqrt{\frac{\pi}{21}}B(3,-1)\\
          -\frac{4\pi}{3\sqrt{10}} B(1,1)
          -\frac{4}{5}\sqrt{\frac{\pi}{21}} B(3,1)]\\
          -\frac{15}{2}\frac{\mu_0 \mathcal{M} e}{2MC}\left[\sum_{r_i=x,y,z} K_{r_i} \int_{K} d\tau O^{r_i}_{R_1 R_1 R_3}\right] \times  [\sqrt{\frac{2 \pi}{105}} B(3,-2)+\frac{2}{5}\sqrt{\frac{\pi}{3}}B(1,0)\\
          -\frac{2}{5}\sqrt{\frac{\pi}{7}}B(3,0)+\sqrt{\frac{2 \pi}{105}}B(3,2)]\\
          -\frac{15}{2}\frac{\mu_0 \mathcal{M} e}{2MC}\left[\sum_{r_i=x,y,z} K_{r_i} \int_{K} d\tau O^{r_i}_{R_2 R_2 R_3}\right] \times 
          [-\sqrt{\frac{2 \pi}{105}} B(3,-2)+\frac{2}{5}\sqrt{\frac{\pi}{3}}B(1,0)\\
          -\frac{2}{5}\sqrt{\frac{\pi}{7}}B(3,0)-\sqrt{\frac{2 \pi}{105}}B(3,2)]\\
          -\frac{15}{2}\frac{\mu_0 \mathcal{M} e}{2MC}\left[\sum_{r_i=x,y,z} K_{r_i} \int_{K} d\tau O^{r_i}_{R_2 R_3 R_3}\right] \times
          [\frac{i}{5}\sqrt{\frac{2 \pi}{3}}B(1,-1)+\frac{4i}{5}\sqrt{\frac{\pi}{21}}B(3,-1)\\
          +\frac{i}{5}\sqrt{\frac{2 \pi}{3}}B(1,1)+\frac{4i}{5}\sqrt{\frac{\pi}{21}}B(3,1)]\\
          -\frac{15}{2}\frac{\mu_0 \mathcal{M} e}{2MC}\left[\sum_{r_i=x,y,z} K_{r_i} \int_{K} d\tau O^{r_i}_{R_1 R_2 R_3}\right] \times
          [i\sqrt{\frac{2 \pi}{105}}B(3,-2)-i\sqrt{\frac{2 \pi}{105}}B(3,2)]
    \end{multline}
    \begin{multline}
        V_{ni}^{(4)}=\frac{3}{2}\frac{\mu_0 \mathcal{M} e}{2MC} \sum_{p=1}^{3}\left\{[\sum_{r_i=x,y,z} K_{r_i} \int _{K} d \tau O^{r_i}_{R_P}] \int \sum_{lm} B(l,m) Y^{*}_{lm}(\theta, \phi) R_{P} d\Omega \right\}\\
        =\frac{3}{2}\frac{\mu_0 \mathcal{M} e}{2MC}\left[\sum_{r_i=x,y,z} K_{r_i} \int _{K} d \tau O^{r_i}_{R_1}\right] \times \sqrt{\frac{2\pi}{3}}[B(1,-1)-B(1,1)]\\
        +\frac{3}{2}\frac{\mu_0 \mathcal{M} e}{2MC}\left[\sum_{r_i=x,y,z} K_{r_i} \int _{K} d \tau O^{r_i}_{R_2}\right] \times 
        i\sqrt{\frac{2\pi}{3}}[B(1,-1)+B(1,1)]\\
        +\frac{3}{2}\frac{\mu_0 \mathcal{M} e}{2MC}\left[\sum_{r_i=x,y,z} K_{r_i} \int _{K} d \tau O^{r_i}_{R_3}\right] \times \sqrt{\frac{4\pi}{3}}B(1,0) 
    \end{multline}
where, $\Vec{K}=\Vec{k}_i+\Vec{k}_n$
\bibliographystyle{plain}
\bibliography{reference}

\end{document}